\documentclass[aps,prc,preprint,amsmath,amssymb,showpacs,preprintnumbers,superscriptaddress]{revtex4-1}
\usepackage{CJKutf8}
\usepackage{graphicx}
\usepackage{dcolumn}
\usepackage{bm}
\usepackage{color}
\usepackage{csquotes}
\usepackage{hyperref}

\allowdisplaybreaks

\begin{document}
\title{Total kinetic energy of fission fragments within time-dependent density functional theory: role of particle-number projection and collective deformations}
\CJKfamily{gbsn}
\author{B. Li }
\affiliation{State Key Laboratory of Nuclear Physics and Technology, School of Physics, Peking University, Beijing 100871, China}
\author{D. Vretenar}
\email{vretenar@phy.hr}
\affiliation{Physics Department, Faculty of Science, University of Zagreb, 10000 Zagreb, Croatia}
\affiliation{State Key Laboratory of Nuclear Physics and Technology, School of Physics, Peking University, Beijing 100871, China}
\author{T. Nik\v si\' c}
\affiliation{Physics Department, Faculty of Science, University of Zagreb, 10000 Zagreb, Croatia}
\author{P. W. Zhao }
\email{pwzhao@pku.edu.cn}
\affiliation{State Key Laboratory of Nuclear Physics and Technology, School of Physics, Peking University, Beijing 100871, China}
\author{J. Meng }
\email{mengj@pku.edu.cn}
\affiliation{State Key Laboratory of Nuclear Physics and Technology, School of Physics, Peking University, Beijing 100871, China}

\begin{abstract}
Time-dependent density functional theory (TDDFT) calculations of fission fragment total kinetic energy (TKE), restricted to axially symmetric trajectories and a single symmetry-violating mean-field state, underestimate TKE by 10-20 MeV away from the most probable mass split. We reexamine this for $^{240}$Pu and $^{252}$Cf within relativistic TDDFT+BCS, combining particle-number projection (PNP) with an extended collective space including hexadecapole and non-axial deformations. Varying the pairing strength leaves the TKE essentially unchanged, but PNP converts each trajectory into a resolved TKE distribution over fragment mass, and the enlarged collective space is needed to reach the correct scission configurations. Together, these reproduce the empirical TKE mass dependence substantially better than standard calculations, indicating the discrepancy reflects missing symmetry restoration and configuration space rather than a deficient energy functional.
\end{abstract}

\date{\today}

\maketitle

\section{Introduction}

Nuclear fission presents one of the richest problems in quantum many-body physics, and a quantitative, microscopic description of the fragments it produces is needed both to test our understanding of the nuclear many-body problem and to support applications ranging from reactor design to the modeling of nucleosynthesis \cite{Schunck2016PPNP,Bender2020,Schunck22PPNP,Cowan2021RMP,Sadhukhan2020PRC,Giuliani2020PRC}. Two quantities are of particular importance: the total kinetic energy (TKE) released between the two nascent fragments, and the excitation energy each fragment carries away from scission.  Because these quantities cannot be measured directly at the extremely short time scales associated with scission, fission-fragment decay codes such as those used for evaluated data rely on largely phenomenological prescriptions - energy-sharing systematics and level-density parametrizations calibrated against the available data \cite{Verbeke2015CPC,Talou2021CPC,Randrup2009PRC,Litaize2010PRC,Becker2013PRC,Litaize2015EPJA,Bulgac2020PRC}. A fully microscopic prediction of TKE and excitation energy, obtained directly from an energy density functional (EDF) without invoking such phenomenology, remains a central goal of fission theory \cite{Schunck2016PPNP,Bender2020,Schunck22PPNP}.

Time-dependent density functional theory (TDDFT), in both its non-relativistic Skyrme, Gogny, and relativistic point-coupling or meson-exchange formulations, has become a practical tool for this purpose \cite{Negele1978PRC,Scamps2015PRC,Bulgac2016PRL,Tanimura2017PRL,Ren2022PRL,Ren2022a,Tong2022PRC,Huang2024PRC,Huang2024EPJA}. In the TDDFT approach to fission, a static, self-consistent solution of the EDF is first constructed on the outer slope of the potential energy surface, beyond the last saddle point; the constraints on the relevant collective (multipole) coordinates are then released and the many-body state is propagated in real time until the two fragments separate at scission. Because pairing correlations control, among other things, the saddle-to-scission time and whether fission occurs at all, essentially all implementations combine the mean-field evolution with a BCS- or Bogoliubov-type treatment of pairing \cite{Scamps2015PRC,Bulgac2016PRL,Bulgac2021PRL,Bulgac2022PRL,Ren2022PRL,Abdurrahman2024PRL,Bulgac2025PRL,Li2026PRL}.

Within this framework, we have previously carried out a systematic study of induced fission of $^{240}$Pu using the relativistic TDDFT+BCS approach built on the point-coupling functional PC-PK1~\cite{Zhao2010PRC}, following self-consistent trajectories in the axial quadrupole-octupole $(\beta_{20},\beta_{30})$ plane from configurations just below the outer barrier to scission. These calculations reproduce the measured TKE reasonably well for the most probable, moderately asymmetric fragmentation, but systematically underpredict it for both more symmetric and highly asymmetric mass splits \cite{Ren2022PRL,Ren2022a,Li2023PRC_FTTD,Li2024Fop,Li2025PRC}. The same qualitative pattern has recently been confirmed, independently and over a far more complete set of fragmentations, by Bjel\v ci\' c, Schunck, and Verriere \cite{Bjelcic2026PRC}, who developed a new axially-symmetric time-dependent Hartree-Fock-Bogoliubov (TDHFB) solver and generated several hundred TDHFB trajectories spanning nearly the entire range of mass splits accessible in the fission of $^{240}$Pu. Their results show that the TKE agrees with experiment only near the most likely fragmentation, while it is underestimated by 10-20 MeV at both small and large mass asymmetries. This discrepancy persists essentially unchanged when the Skyrme parametrization, the pairing strength, or the surface-versus-volume character of the pairing interaction are varied. Taken together, these results from two independent, complementary microscopic approaches (relativistic TDDFT+BCS in our earlier work and non-relativistic Skyrme TDHFB in Ref. \cite{Bjelcic2026PRC}) indicate that the disagreement with data reflects a shared limitation of standard (TD)DFT-based fission dynamics, rather than a deficiency of any one energy functional or numerical implementation.

Two ingredients that are typically absent from this standard treatment could plausibly account for at least part of the discrepancy. The first concerns particle-number symmetry: once pairing correlations are included, the many-body state is no longer an eigenstate of the neutron and proton number operators, and spontaneously breaks the U(1) gauge symmetry associated with particle-number conservation. As a result, the fission fragments extracted from a single TDDFT+BCS or TDHFB trajectory are characterized only by average, generally non-integer, proton and neutron numbers, and the TKE computed along the trajectory refers to this symmetry-violating, averaged configuration rather than to fragments with sharp, physical (Z, N). Restoring the symmetry by particle-number projection (PNP) \cite{Scamps2013PRC} makes it possible to extract, from a single trajectory, the distribution of TKE values associated with the individual channels of good particle number the trajectory actually populates, and to compare this distribution directly with the measured TKE as a function of fragment mass, rather than with a single average point. The second ingredient is the collective space explored on the way to scission: most TDDFT studies of fission, including our own earlier work, are restricted to trajectories confined to the axially symmetric $(\beta_{20},\beta_{30})$ plane, even though reaching scission for near-symmetric configurations, or reproducing the correct mass split for very asymmetric ones, may require additional collective coordinates such as the hexadecapole deformation $\beta_{40}$, which controls the thickness of the pre-scission neck, or non-axial degrees of freedom such as the triaxial $\beta_{22}$  and asymmetric-octupole $\beta_{31}$  parameters.

In this work we reexamine the total kinetic energy of fission fragments of $^{240}$Pu and $^{252}$Cf within the relativistic TDDFT+BCS framework, combining particle-number projection with an extended set of collective coordinates beyond the standard $(\beta_{20},\beta_{30})$ deformation parameters. Section \ref{sec_theo} summarizes the TDDFT+BCS formalism and the particle-number projection method used to compute the TKE of fragments with fixed proton and neutron numbers. Section \ref{sec_num} describes the numerical implementation. In Sec. \ref{TKE} we present results for $^{240}$Pu and $^{252}$Cf, compare the TKE obtained with and without particle-number projection along several representative fission paths, and quantify the effect of enlarging the collective space on the resulting TKE(A) systematics. Section \ref{sec_sum} summarizes our findings and discusses their implications for microscopic fission theory.

\section{Theoretical  framework}\label{sec_theo}

\subsection{Time-dependent density functional theory}

In the time-dependent BCS approximation of TDRDFT~\cite{Ebata2010PRC,Scamps2013PRC}, the nuclear wavefunction reads
\begin{equation}
   |\Phi (t)\rangle = \prod_{k>0}[\mu_{k}(t)+\nu_{k}(t)c_{k}^\dagger(t)c_{\bar{k}}^\dagger(t)]|-\rangle \;.
\label{Eq_Slater}
\end{equation}
In Eq.~(\ref{Eq_Slater}), $\mu_{k}(t)$ and $\nu_{k}(t)$ are the parameters of the transformation between the canonical and quasiparticle bases,
and $c_{k}^\dagger(t)$ denotes the creation operator associated with the canonical state $\phi_k(\bm{r},t)$.
~The time evolution of $\phi_k(\bm{r},t)$ is determined by the time-dependent Dirac equation~\cite{Ren2022PRL,Ren2022a}
\begin{equation}\label{Eq_td_Dirac_eq_BCS}
  i\frac{\partial}{\partial t}\phi_k(\bm{r},t)=[\hat{h}(\bm{r},t)-\varepsilon_k(t)]\phi_k(\bm{r},t),
\end{equation}
where $\varepsilon_k(t)=\langle\phi_k(\bm{r},t)|\hat{h}(\bm{r},t)|\phi_k(\bm{r},t)\rangle$, and the single-particle Hamiltonian $\hat{h}(\bm{r},t)$ is determined self-consistently
at each step in time by the time-dependent densities and currents in the scalar, vector, and isovector channels,
\begin{subequations}\label{Eq_density_current}
  \begin{align}
    &\rho_S(\bm{r},t)=\sum_k n_{k}(t)\bar{\phi}_k(\bm{r},t)\phi_k(\bm{r},t),\\
    &j^{\mu}(\bm{r},t)=\sum_k n_{k}(t)\bar{\phi}_k(\bm{r},t)\gamma^\mu\phi_k(\bm{r},t),\\
    &j_{TV}^{\mu}(\bm{r},t)=\sum_k  n_{k}(t)\bar{\phi}_k(\bm{r},t)\gamma^\mu\tau_3\phi_k(\bm{r},t),
  \end{align}
\end{subequations}
where $\tau_3$ is the isospin Pauli matrix, and the vector density reads $\rho_V(\bm{r},t) = j^0(\bm{r},t)$. The time evolution of the occupation probability $n_{k}(t)=|\nu_{k}(t)|^2$,
and pairing tensor $\kappa_{k}(t)=\mu_{k}^{*}(t)\nu_{k}(t)$, is governed by the following equations:
\begin{subequations}\label{Eq_td_nkapp_eq_BCS}
   \begin{align}
     &i\frac{d}{dt}n_{k}(t)=\kappa_{k}(t)\Delta_{k}^*(t)-\kappa_{k}^{*}(t)\Delta_{k}(t),\\
     &i\frac{d}{dt}\kappa_{k}(t)=[\varepsilon_k(t)+\varepsilon_{\bar{k}}(t)]\kappa_{k}(t)+\Delta_{k}(t)[2n_{k}(t)-1] ,
   \end{align}
\end{subequations}
In the BCS approximation a monopole pairing interaction is employed, and the gap parameter $\Delta_{k}(t)$ is defined as,
\begin{equation}
  \Delta_{k}(t)=\left[G\sum_{k'>0}f(\varepsilon_{k'})\kappa_{k'}\right]f(\varepsilon_k),
\end{equation}
where $f(\varepsilon_k)$ is the cut-off function for the pairing window~\cite{Scamps2013PRC}, and $G$ is the pairing strength.

\subsection{Total kinetic energy of fission fragments}

For a TDDFT trajectory, the average total kinetic energy (TKE) of the two fission fragments reads
\begin{equation}
\bar{E}_{TKE}=\frac{1}{2}m\bar{A}_{\rm H}\bm{v}^2_{{\rm H}}+\frac{1}{2}m\bar{A}_{\rm L}\bm{v}^2_{{\rm L}}+E_{\rm Coul},
\label{Eq_ave_TKE}
\end{equation}
where $\bm{v}_{f}$ is the velocity of the fragment $f={\rm H,L}$
\begin{equation}
\bm{v}_{f}=\frac{1}{\bar{A}_{f}}~\int_{V_f} d\bm{r}~\bm{j}(\bm{r}) \;,
\end{equation}
$\bm{j}(\bm{r})$ is the current, and $\bar{A}_f$ is the average mass number of fragment $f$.
The integration is over the half-volume corresponding to fragment $f$. $E_{\mathrm{Coul}}$ is the Coulomb energy,
\begin{equation}
E_{\rm Coul} = \alpha \int_{V_H}\int_{V_L} \frac{\rho_V^{p}({\bm r_H})\rho_V^{p}({\bm r_L})}{|{\bm r_H}-{\bm r_L}|} d{\bm r_H} d{\bm r_L},
\end{equation}
where ${\bm r_H}$ (${\bm r_L}$) is the space coordinate in the subspace $V_H$ ($V_L$), and $\rho_V^{p}({\bm r})$ is the proton vector density.

When calculating $\bar{E}_{TKE}$ on a TDDFT+BCS trajectory, the states of the fission fragments are not eigenstates of the nucleon number operator.
To restore gauge symmetry we employ particle number projection, and the corresponding TKE of fission fragments with good neutron and proton number reads
\begin{equation}
E_{\rm TKE} (Z_{\rm L},N_{\rm L},Z_{\rm H},N_{\rm H})=
\frac{1}{2}mA_{\rm H}\bm{v}^2_{{\rm H}} (Z_{\rm H},N_{\rm H}) +\frac{1}{2}mA_{\rm L}\bm{v}^2_{{\rm L}}(Z_{\rm L},N_{\rm L}) +E_{\rm Coul}(Z_{\rm L},N_{\rm L},Z_{\rm H},N_{\rm H}).
\label{Eq_TKE}
\end{equation}
$Z_{\rm L}$, $N_{\rm L}$, and $A_{\rm L}$ ($Z_{\rm H}$, $N_{\rm H}$, and $A_{\rm H}$) are the proton, neutron, and mass number of the light (heavy) fragment.
They need to satisfy,
\begin{equation}
Z = Z_L + Z_H,~~N = N_L + N_H,~~ A = A_L+A_H,
\end{equation}
where $Z$, $N$, and $A$ are the proton, neutron, and mass number of the total fissioning system.
The velocity of the fragment $f={\rm H,L}$ with proton number $Z_f$ and neutron number $N_f$ reads
\begin{equation}
\bm{v}_{f} (Z_{f},N_{f}) = \frac{1}{A_{f}}~\int_{V_f} d\bm{r}~\bm{j}(\bm{r},Z_{f},N_{f}) \;.
\end{equation}
$E_{\rm Coul}(Z_{\rm L},N_{\rm L},Z_{\rm H},N_{\rm H})$ is the Coulomb energy,
\begin{equation}
E_{\rm Coul}(Z_{\rm L},N_{\rm L},Z_{\rm H},N_{\rm H}) = \alpha \int_{V_H}\int_{V_L} \frac{\rho_V^{p}({\bm r_H},Z_{\rm H},N_{\rm H})\rho_V^{p}({\bm r_L},Z_{\rm L},N_{\rm L})}{|{\bm r_H}-{\bm r_L}|} d{\bm r_H} d{\bm r_L},
\end{equation}
where $\alpha$ is the fine structure constant. The projected density and current read
\begin{equation}
    j^\mu({\bm r_f},Z_f,N_f) =
    \frac{\langle \Phi |\bar{\psi} \gamma^\mu \psi \hat{P}_N\hat{P}_{N_f}^{V_f}\hat{P}_Z\hat{P}_{Z_f}^{V_f}|\Phi\rangle}
    {\langle\Phi|\hat{P}_N\hat{P}_{N_f}^{V_f}\hat{P}_Z\hat{P}_{Z_f}^{V_f}|\Phi\rangle}.
    \label{Eq_pj_curr}
\end{equation}
where $\psi$ is the nucleon field operator.
$\hat{P}_{M_f}^{V_f}$ and $\hat{P}_M$ ($M = Z,N$) are the particle number projection operators defined in the subspace $V_f$ and the entire space \cite{Scamps2013PRC}, respectively,
\begin{equation}
    \hat{P}_{M_f}^{V_f}=\frac{1}{2\pi}\int_0^{2\pi} d\theta_f~e^{-i\theta_f(M_f-\hat{M}_f)},
    \label{P_n_2}
\end{equation}
\begin{equation}
    \hat{P}_M=\frac{1}{2\pi}\int_0^{2\pi} d\theta~e^{-i\theta(M-\hat{M})},
    \label{P_N}
\end{equation}
where $\hat{M}_f$ is the particle number operator in the subspace $V_f$, and $\hat{M}$~is the particle number operator in the entire space.
The detailed expressions for proton and neutron number projection are included in the Appendix~\ref{sec_app}.\
The TKE of the fission fragments as a function of heavy-fragment mass number $A_H$ is given by,
\begin{equation}
    E_{\rm TKE}(A_H) = \frac{\sum\limits_{\substack{Z_H,N_H,Z_H+N_H=A_H}} E_{\rm TKE}(Z_{\rm H},N_{\rm H})P(Z_{\rm H},N_{\rm H})}{\sum\limits_{\substack{Z_H,N_H,Z_H+N_H=A_H}} P(Z_{\rm H},N_{\rm H})}.
    \label{Eq_TKE_A}
\end{equation}
Here $E_{\rm TKE}(Z_{\rm H},N_{\rm H})$ denotes $E_{\rm TKE}(Z-Z_{\rm H},N-N_{\rm H},Z_{\rm H},N_{\rm H})$ in Eq.~(\ref{Eq_TKE}),
since the light-fragment particle numbers are fixed by $N_L = N-N_H$ and $Z_L = Z-Z_H$.
The probability of obtaining a heavy fragment with neutron number $N_{\rm H}$ and proton number $Z_{\rm H}$,
conditional on the total fissioning system having neutron number $N$ and proton number $Z$, is
\begin{equation}
P(Z_{\rm H},N_{\rm H}) = \frac{\langle\Phi|\hat{P}_N\hat{P}_{N_H}^{V_H}\hat{P}_Z\hat{P}_{Z_H}^{V_H}|\Phi\rangle}{\langle\Phi|\hat{P}_N\hat{P}_Z|\Phi\rangle}.
\end{equation}

\section{Numerical details}\label{sec_num}
In the present calculation that employs the time-dependent relativistic density functional theory in 3D-coordinate space \cite{Ren2017PRC,Ren2019SCPMA,Ren2020NPA,Li2020PRC,Xu2024PRC,Xu2024PLB,Xu2024PRL,Xu2026PRL},
the mesh spacing of the lattice is 1.0 fm for all directions.
The box size is $L_x\times L_y\times L_z=20\times20\times60~{\rm fm}^3$ for $^{240}$Pu, and
$L_x\times L_y\times L_z=26\times26\times60~{\rm fm}^3$ for $^{252}$Cf.
The time-dependent Dirac equation \eqref{Eq_td_Dirac_eq_BCS} is solved using the predictor-corrector method, and the time-dependent equations \eqref{Eq_td_nkapp_eq_BCS} with the Euler algorithm.
The step for the time evolution is $0.2~{\rm fm}/c~\approx 6.67\times10^{-4}$~zs.
The point-coupling relativistic energy density functional PC-PK1~\cite{Zhao2010PRC} is used in the particle-hole channel, together with a monopole pairing interaction in the particle-particle channel.
The pairing strength parameters $G$ are -0.135 MeV for neutrons and -0.230 MeV for protons in the case of $^{240}$Pu,
and -0.125 MeV (neutrons) and -0.210 MeV (protons)  for $^{252}$Cf.
These values are determined by the empirical pairing gaps of $^{240}$Pu and $^{252}$Cf, respectively, using the three-point odd-even mass formula.
The subspace $V_f$ is selected as $z>0$ or $z<0$ , with the $z$-axis along the fission direction, and the interval of $z$ is from $-30$~fm to $30$~fm.
In the computation of~Eq.~(\ref{Eq_pj_curr}), the integrals over $\theta$ and $\theta_f$ are evaluated with the trapezoidal rule, and  the interval is discretized with 40 points.
Only channels with $P(Z_{\rm H},N_{\rm H})>10^{-6}$ are included in the sums in Eq.~(\ref{Eq_TKE_A}).
\section{Results and discussion}\label{TKE}
\begin{figure}[!htbp]
\centering
\includegraphics[width=0.5\textwidth]{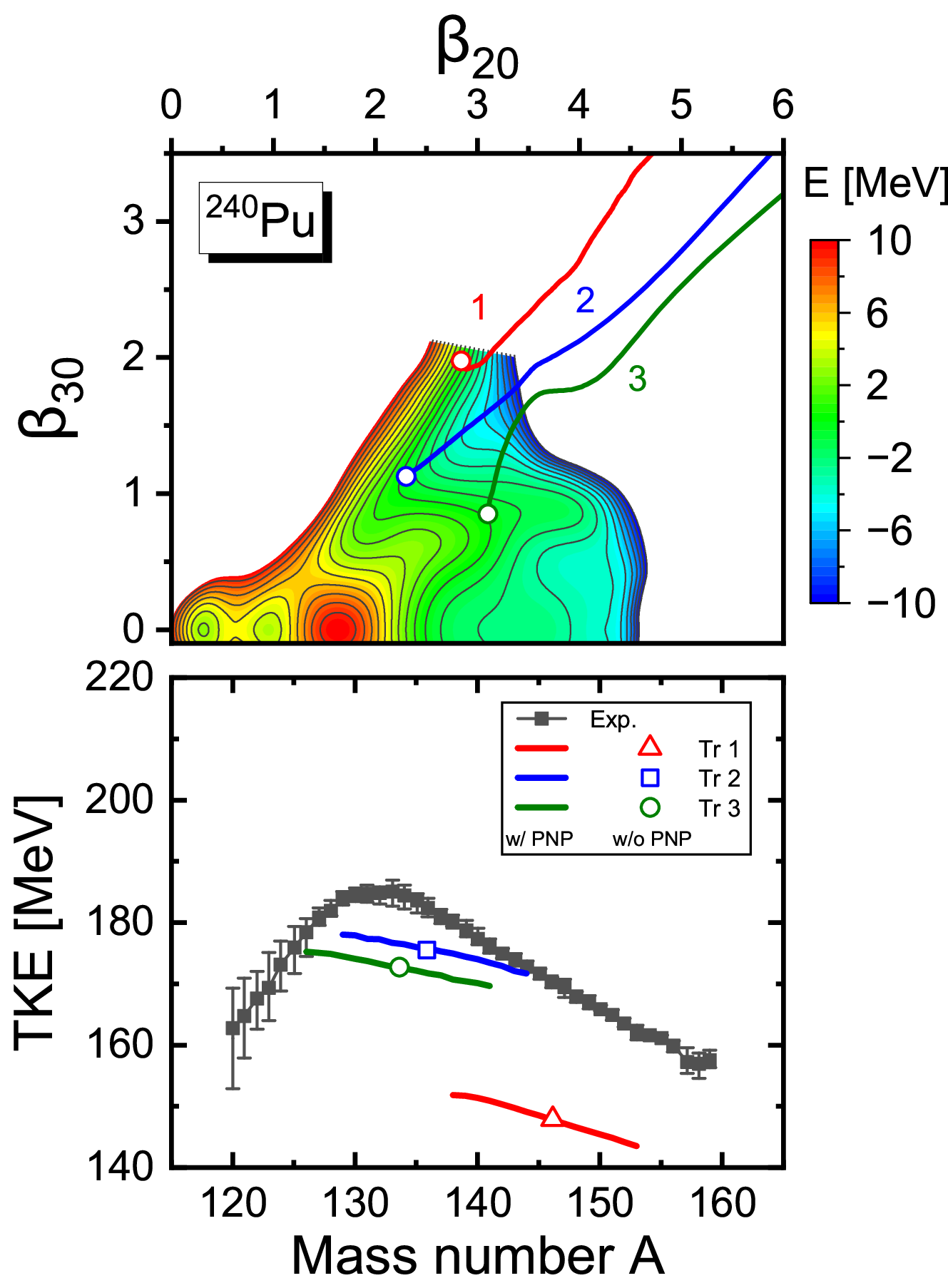}
\caption{(Color online) Top panel: Self-consistent deformation-energy surface of $^{240}$Pu in the plane of axially symmetric quadrupole $\beta_{20}$ and octupole $\beta_{30}$ deformation parameters. Contours join points on the surface with the same energy, and the open dots are on the isoenergy curve at 1 MeV below the energy of the equilibrium minimum.
The corresponding deformation parameters are $(\beta_{20},\beta_{30})= (2.84,1.98), (2.30,1.13), (3.10,0.85)$, and these are the starting points for the self-consistent TDDFT fission trajectories 1, 2 and 3. Bottom panel: Open symbols denote the average TKE of the emerging fragments from trajectories $1-3$. The curves are the corresponding particle-number projected TKE distributions of these trajectories, as functions of the heavy fragment mass. Experimental data are obtained by averaging results from Refs.~\cite{Wagemans1984PRC,Geltenbort1986RE,NISHIO1995JNST,TSUCHIYA2000JNST}.
}
\label{fig_1}
\end{figure}

The deformation-energy surface of $^{240}$Pu, calculated with the relativistic density functional PC-PK1~\cite{Zhao2010PRC} plus monopole pairing, is shown in the top panel of Fig.~\ref{fig_1} as a function of the axially symmetric quadrupole $\beta_{20}$ and octupole $\beta_{30}$ deformation parameters.
The equilibrium minimum is at $(\beta_{20},\beta_{30})=(0.29,0.0)$, and the isomer is located at $(\beta_{20},\beta_{30})=(0.94,0.0)$.
There are two fission barriers: the inner one at $(\beta_{20},\beta_{30}) \approx (0.63,0.0)$, and the outer barrier at $(\beta_{20},\beta_{30}) \approx (1.27,0.36)$.

Similar to our previous studies~\cite{Ren2022PRL,Ren2022a,Li2023PRC_FTTD,Li2025PRC}, we start with three points on the energy surface, located 1 MeV below the energy of the equilibrium minimum and denoted by open dots in the top panel of Fig.~\ref{fig_1}. Upon releasing the constraints on the quadrupole and octupole moments, the system evolves toward scission along the corresponding self-consistent TDDFT trajectories. Following scission, two fission fragments (FFs) emerge with total kinetic energy (TKE). The time evolution is terminated when the distance between the centers of mass of the two fission fragments reaches 25 fm. The average mass number and total kinetic energy of the fission fragments are then evaluated using the vector density $\rho_V({\bm r})$ and Eq.~(\ref{Eq_ave_TKE}), respectively. The corresponding results for trajectories $1-3$ are shown as open symbols in the bottom panel of Fig.~\ref{fig_1}, together with the experimental values. While the calculated TKE for trajectories 2 and 3 are in qualitative agreement with the data, the TKE predicted for the highly mass-asymmetric fission path 1 is more than 20 MeV lower than the experimental value.

Because the inclusion of pairing correlations in the self-consistent calculation of deformation-energy surfaces and fission paths breaks the U(1) particle-number gauge symmetry, the calculated fragment mass numbers are expectation values and, in general, are not restricted to integers. Particle-number symmetry can be restored by particle-number projection (PNP). To obtain states with well-defined particle numbers for both the individual fission fragments (FFs) and the total system, we employ the double-particle-number projection method~\cite{Scamps2013PRC}. For each trajectory, we therefore calculate the TKEs associated with the mass-number distributions obtained from PNP. The resulting TKEs are shown in the lower panel of Fig.~\ref{fig_1}. For each trajectory, the particle-number projected TKE decreases with increasing heavy-fragment mass number, primarily because of the decreasing Coulomb repulsion between the two fragments.

Thus, the standard TDDFT approach provides the TKE only for fission fragments characterized by their average mass numbers. With the inclusion of PNP, however, it becomes possible to calculate the TKE for a range of fragment mass numbers centered around the average values. The resulting TKE distributions reproduce the experimentally observed mass dependence. As an example, for trajectory 2, PNP enables us to calculate the TKE for heavy-fragment mass numbers ranging from 129 to 144, with the corresponding TKE decreasing from 178.0 MeV to 171.7 MeV.

\begin{figure}[!htbp]
\centering
\includegraphics[width=0.5\textwidth]{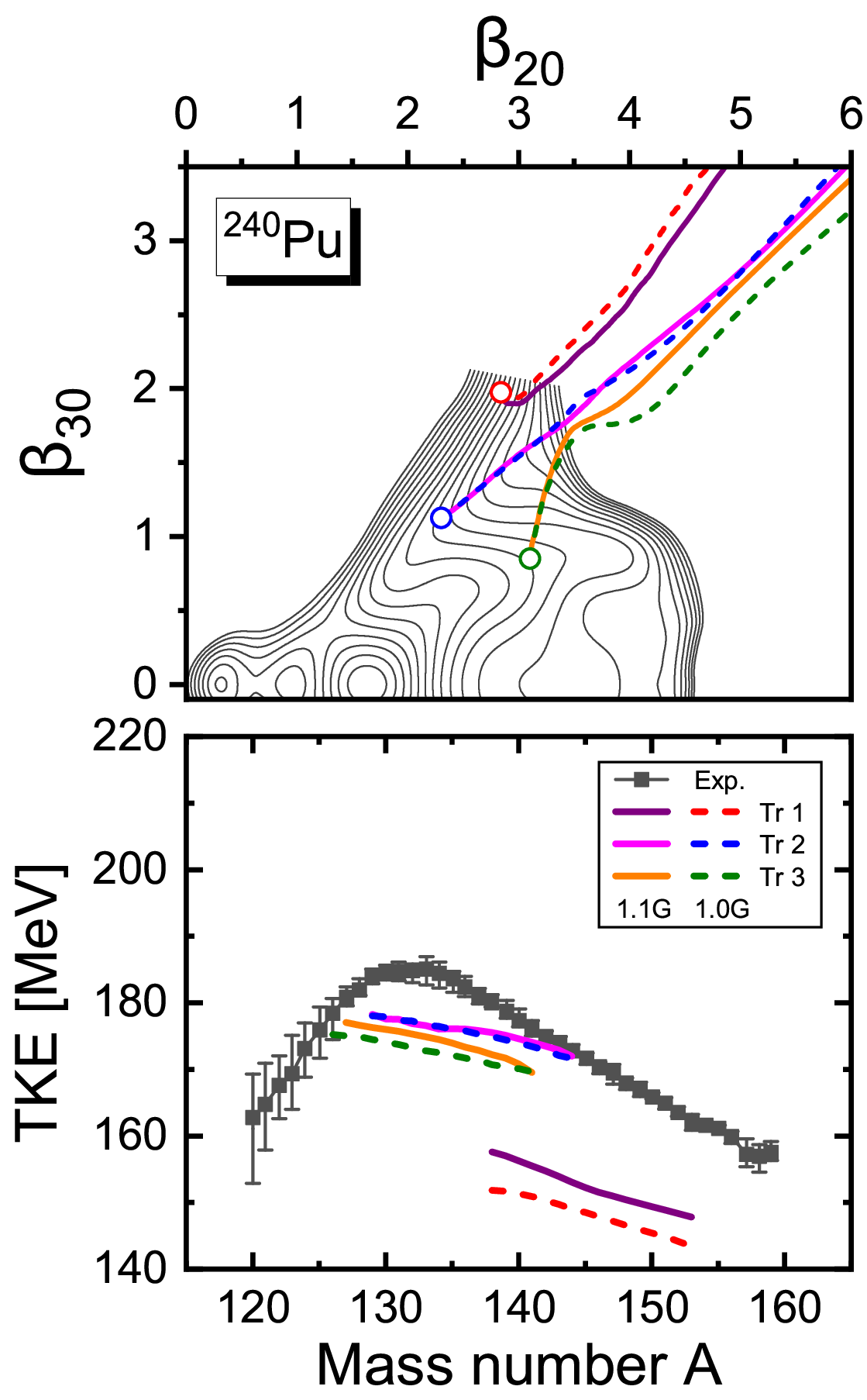}
\caption{(Color online) Same as in the caption to Fig.~\ref{fig_1}, but also including the TKE distributions for three
additional TDDFT trajectories obtained by increasing the pairing strength parameters by $10\%$.
}
\label{fig_2}
\end{figure}

In Ref.~\cite{Bjelcic2026PRC} the effect of pairing correlations on the calculated TKEs was investigated by varying both the strength of the pairing interaction and its surface-versus-volume character. It was found that none of these modifications significantly reduces the discrepancy with experiment: the underestimation of the TKE away from the most probable fission channel persists essentially unchanged. To further test this conclusion using a different pairing interaction, we have recalculated the self-consistent fission paths starting from the same initial points as those shown in Fig.~\ref{fig_1}, but with the pairing-strength parameter increased by 10\%.
We note that the original pairing strengths were determined from the empirical pairing gaps of $^{240}$Pu. The results are compared in Fig.~\ref{fig_2}. The fission paths are only marginally affected, and the corresponding TKEs show only a slight improvement compared with the experimental data.

\begin{figure}[!htbp]
\centering
\includegraphics[width=0.5\textwidth]{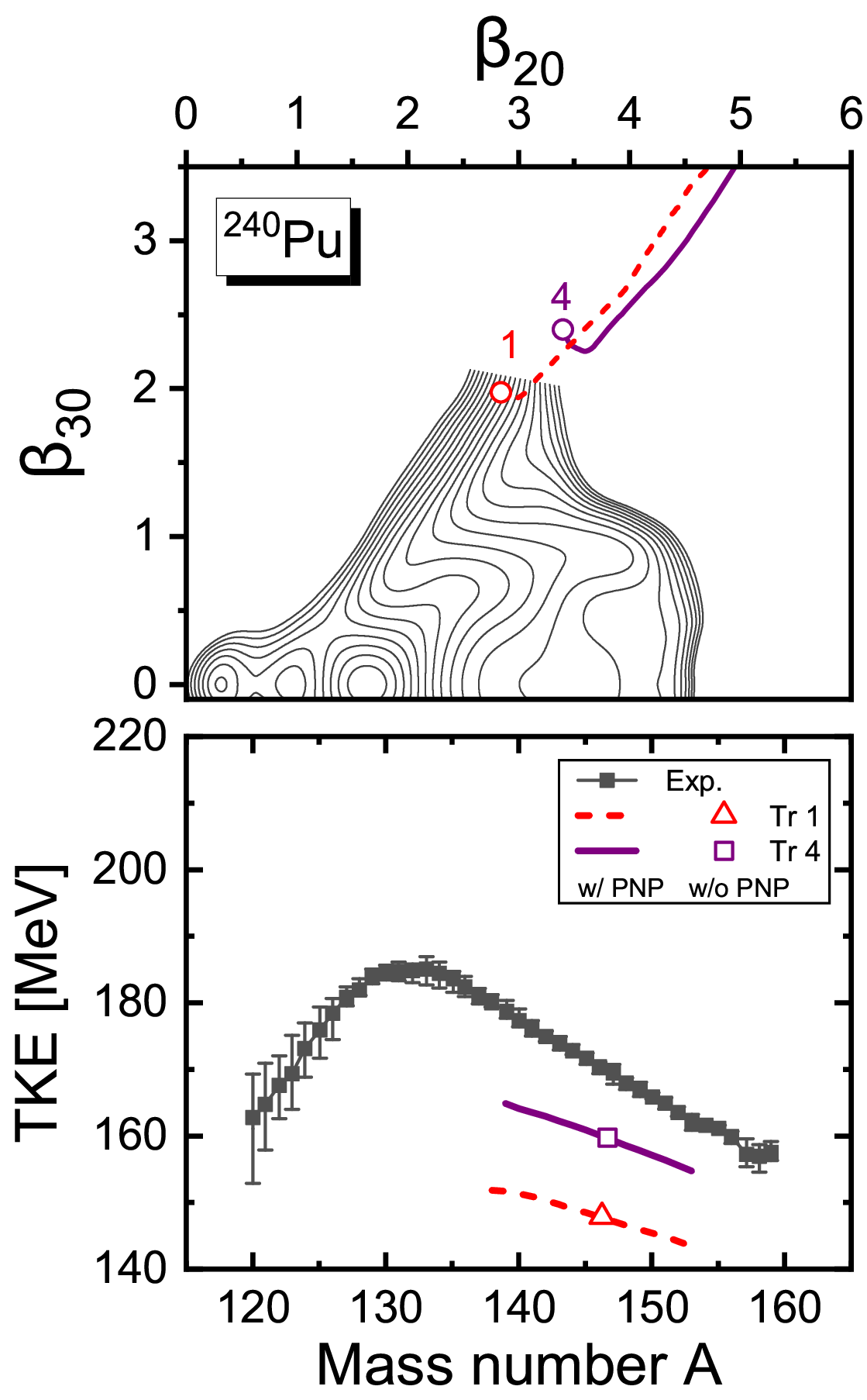}
\caption{(Color online) Top: The deformation energy surface and trajectory 1 are the same as in Fig.~\ref{fig_1}.
Trajectory 4, shown as the solid curve, starts at $(\beta_{20},\beta_{30},\beta_{40})= (3.40,2.40,8.20)$ and includes an initial $\beta_{40}$ constraint. It starts at the same energy as trajectory 1.
Bottom: The results for trajectory 1 and the data are the same as in Fig.~\ref{fig_1}.
The open square denotes the average TKE of the fission fragments for trajectory 4, and the solid curve is the corresponding particle-number projected TKE distribution.
}
\label{fig_3}
\end{figure}

\begin{figure}[!htbp]
\centering
\includegraphics[width=0.8\textwidth]{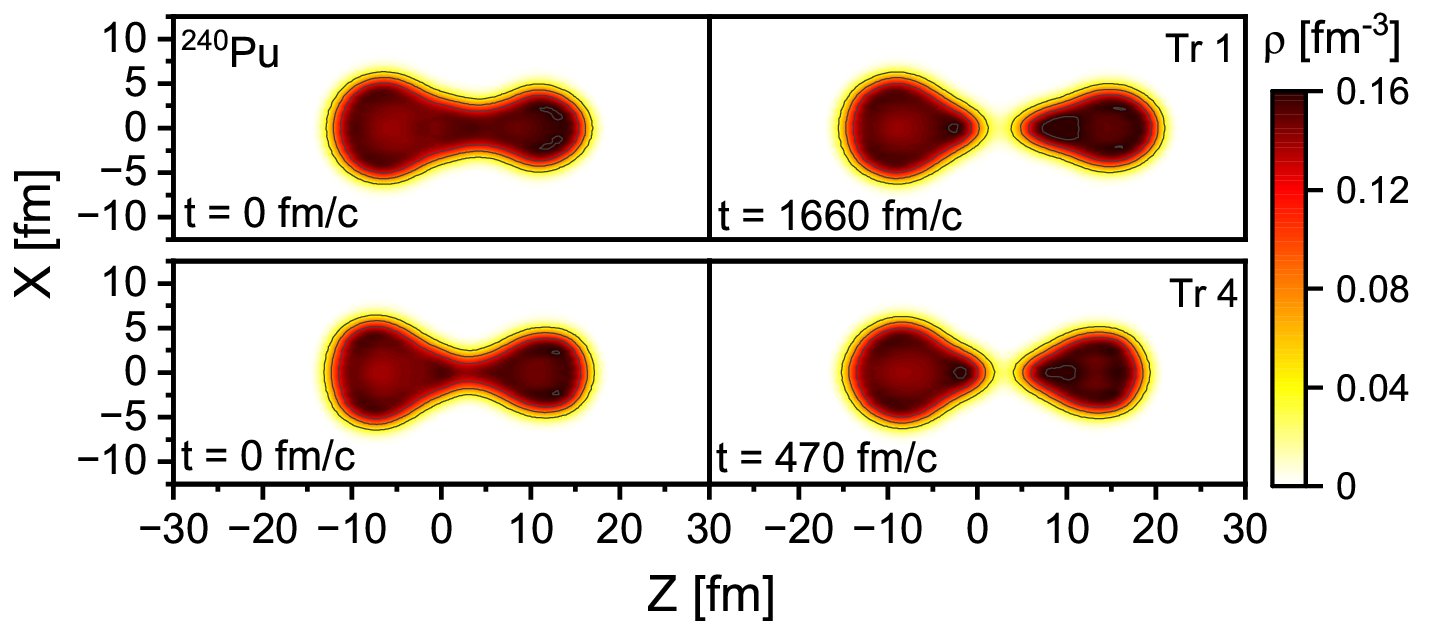}
\caption{(Color online) The one-body density of $^{240}$Pu at the initial (left) and scission points (right), for fission trajectories 1 (top) and 4 (bottom), shown in Fig.~\ref{fig_3}.
}
\label{fig_4}
\end{figure}

Most applications of TDDFT to fission dynamics have assumed axial symmetry, and the majority of calculations of fission yields and TKEs have been restricted to self-consistent trajectories on the quadrupole-octupole $(\beta_{20},\beta_{30})$ plane. In the present analysis, we begin to explore additional collective degrees of freedom and their effects on the predicted TKEs. Figure~\ref{fig_3} compares the mass-asymmetric fission path 1 with the projection on the $(\beta_{20},\beta_{30})$ plane of trajectory 4, which starts at $(\beta_{20},\beta_{30},\beta_{40})= (3.40,2.40,8.20)$ and includes an initial $\beta_{40}$ constraint. Although trajectory 4 starts at the same energy as trajectory 1, constraining the initial hexadecapole deformation and, consequently, the initial neck thickness of the fissioning system leads to a smaller separation between the charge distributions of the FFs at scission. This is illustrated in Fig.~\ref{fig_4}, which compares the nucleon density of $^{240}$Pu at the initial and scission points for fission trajectories 1 and 4. As a result, the Coulomb repulsion at scission is stronger for trajectory 4. Thus, despite the similar fragment mass distribution obtained in the two cases, the corresponding TKEs differ by almost 10 MeV.

\begin{figure}[!htbp]
\centering
\includegraphics[width=0.5\textwidth]{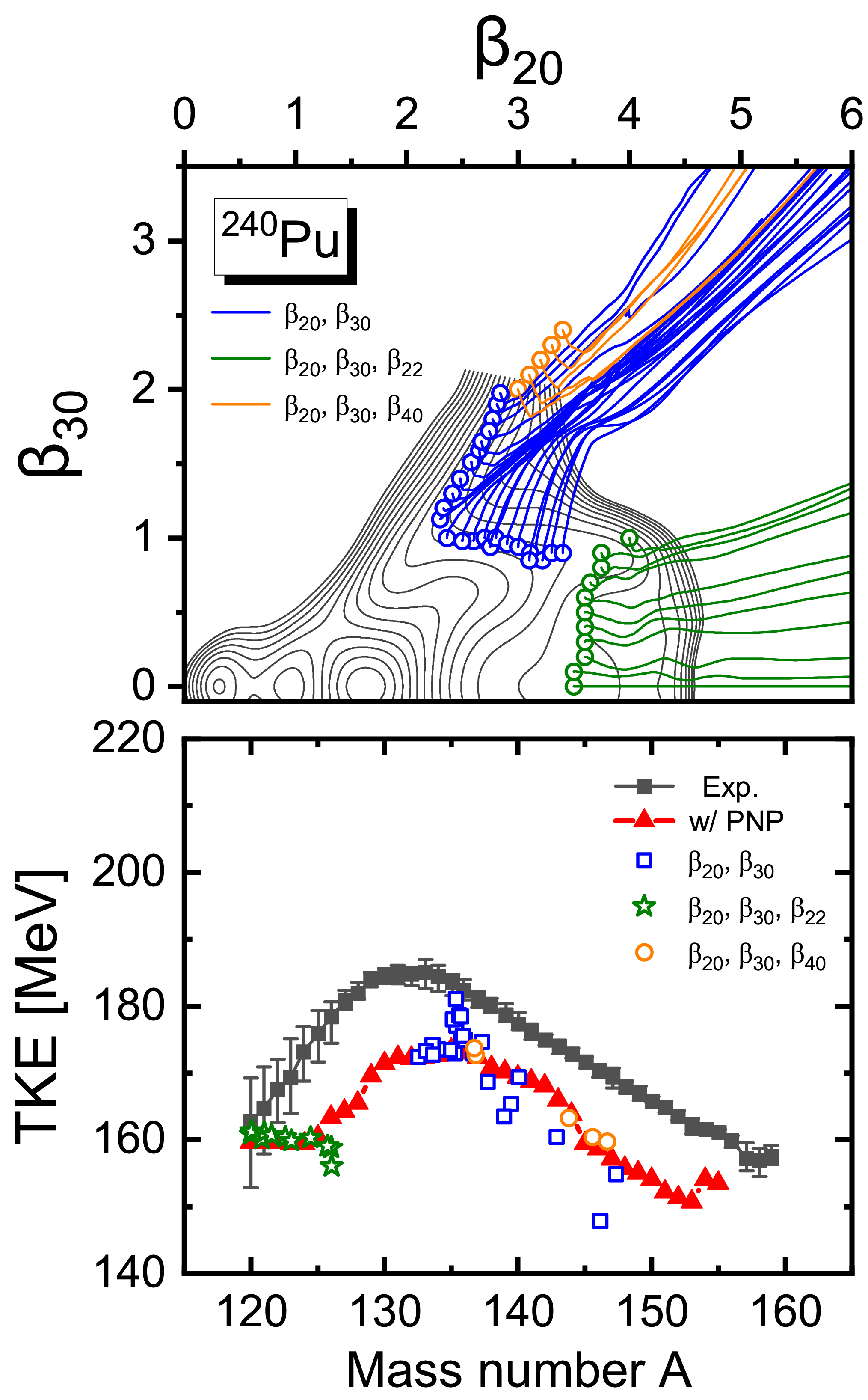}
\caption{(Color online) Top: The deformation energy surface is the same as in the top panel of Fig.~\ref{fig_1}.
The open dots denote the initial $\beta_{20}$ and $\beta_{30}$ deformation parameters for the TDDFT fission trajectories, with all initial configurations having the same energy, 1 MeV below that of the equilibrium minimum.
The solid curves represent the projections of these trajectories onto the ($\beta_{20}, \beta_{30}$) plane.
The blue, green, and orange TDDFT trajectories start from initial states constrained with
$(\beta_{20},\beta_{30})$, $(\beta_{20},\beta_{30}, \beta_{22})$, and $(\beta_{20},\beta_{30},\beta_{40})$ deformation parameters, respectively.
Bottom: The blue, green, and orange symbols show the average TKE of the emerging fragments
from TDDFT trajectories starting at the corresponding initial points shown in the top panel.
The red triangles show the average of the particle-number projected TKE distributions obtained from all TDDFT fission trajectories in the top panel.
}
\label{fig_5}
\end{figure}

This result highlights the importance of exploring a broader space of collective degrees of freedom when modeling fission observables. In Fig.~\ref{fig_5} we present the results of a calculation in which a substantially larger number of trajectories has been considered, with all initial configurations having the same energy, 1 MeV below that of the equilibrium minimum. The upper panel shows the trajectories projected onto the $(\beta_{20}, \beta_{30}$) plane. The blue trajectories start from initial states constrained in $(\beta_{20},\beta_{30})$ and lie within the principal fission valley. The more asymmetric trajectories (orange) start from states constrained with the  initial $(\beta_{20},\beta_{30},\beta_{40})$ deformation parameters. Finally, to reach the scission contour in the mass-symmetric region, axial symmetry has to be broken, and the green trajectories are initiated from points constrained by  $(\beta_{20},\beta_{30}, \beta_{22})$. Axially symmetric trajectories in the mass-symmetric region do not reach the scission contour.

The resulting average TKEs are shown in the lower panel of Fig.~\ref{fig_5} as open symbols, using the same color coding. Once again, only the calculated TKEs for the most probable fission paths in the asymmetric valley are relatively close to the experimental values. For the more symmetric and highly asymmetric trajectories, neither the absolute values nor the mass dependence of the calculated TKEs reproduce the experimental data.

The red triangles show the TKEs obtained by averaging over the PNP distributions. In this procedure, several trajectories can contribute to a given pair of fission fragments. For each trajectory, the PNP distribution is first calculated, and the TKE associated with a particular fragment pair is then obtained by averaging over all contributions to that pair.
\begin{equation}
    \langle E_{\rm TKE} (A_H)\rangle = \frac{1}{m}\sum_{q=1}^{m} E^{q}_{\rm TKE} (A_H),
\end{equation}
where $E^{q}_{\rm TKE} (A_H)$ is the TKE for the $q$-th TDDFT trajectory, calculated using Eq.~(\ref{Eq_TKE_A}),
and $m$ is the number of TDDFT trajectories contributing to the TKE at a given heavy-fragment mass number $A_H$.
Although the calculated TKEs still do not reproduce the experimental values quantitatively, their mass dependence is now in substantially better agreement with the data.

The TKEs obtained by averaging over the PNP distributions, shown in Fig.~\ref{fig_5} , should be regarded as lower bounds on the total kinetic energy. This is because the excitation energy of the fissioning system is not included: all trajectories are initialized at points on the deformation-energy surface of the cold nucleus. The calculated TKEs are compared with experimental data for the mean kinetic-energy release in
fission of the compound nucleus $^{240}$Pu \cite{Wagemans1984PRC,Geltenbort1986RE,NISHIO1995JNST,TSUCHIYA2000JNST}.
In these experiments, the compound nucleus is formed by thermal-neutron capture on $^{239}$Pu, with an excitation energy corresponding to the neutron separation energy.
However, this pre-scission excitation energy cannot simply be added to the TKE, since a fraction of it is converted into the excitation energy of the nascent fragments.

\begin{figure}[!htbp]
\centering
\includegraphics[width=0.5\textwidth]{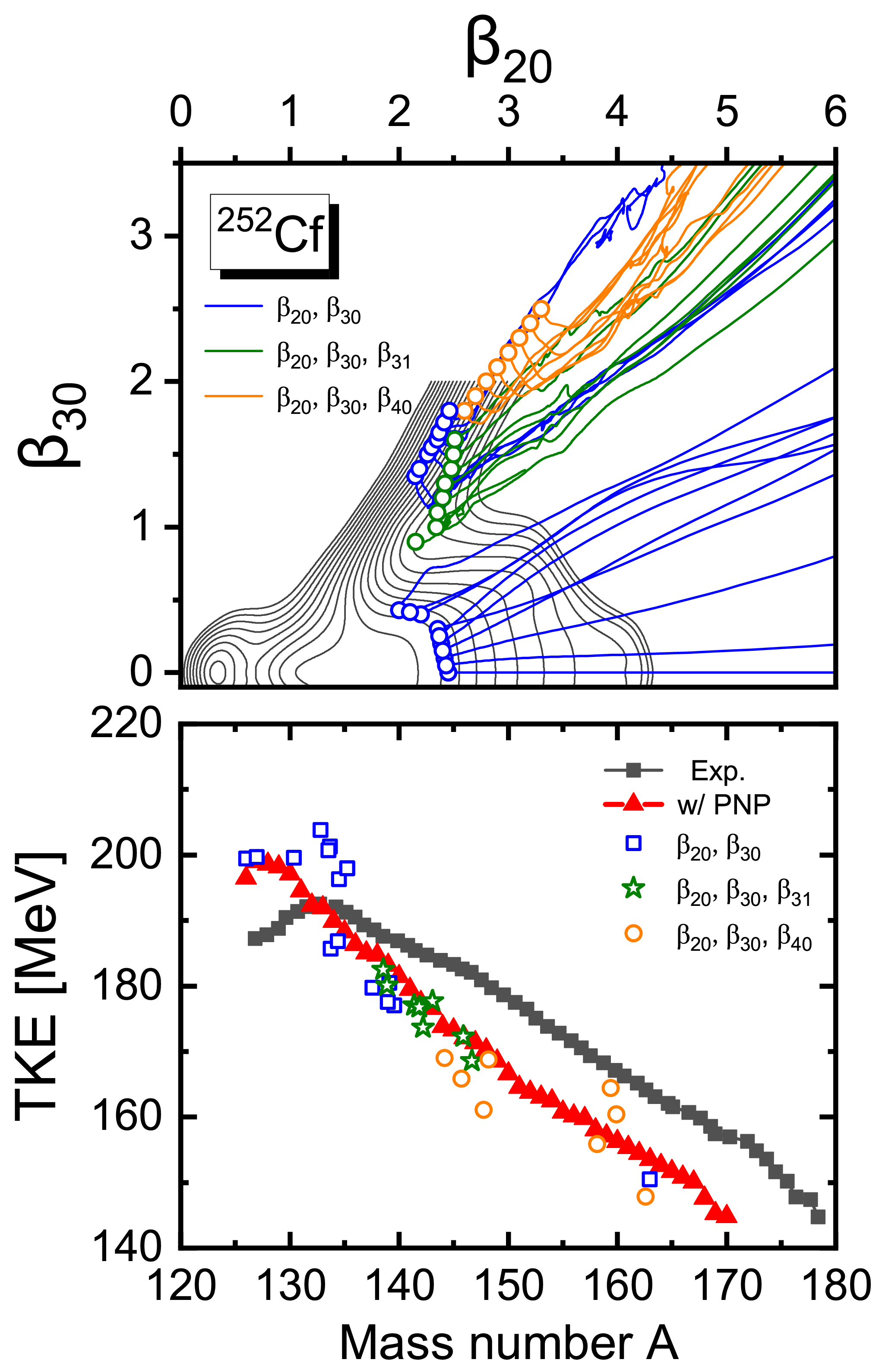}
\caption{(Color online) Top: Self-consistent deformation energy surface of $^{252}$Cf in the plane of axially symmetric quadrupole $\beta_{20}$ and octupole $\beta_{30}$ deformation parameters.
The open dots denote the initial $\beta_{20}$ and $\beta_{30}$ deformation parameters for the TDDFT fission trajectories, with all initial configurations having the same energy as the equilibrium minimum.
The solid curves represent the projections of these trajectories onto the ($\beta_{20}, \beta_{30}$) plane. The
blue, green, and orange TDDFT trajectories start from initial states constrained with $(\beta_{20},\beta_{30})$, $(\beta_{20},\beta_{30}, \beta_{31})$, $(\beta_{20},\beta_{30},\beta_{40})$ deformation parameters, respectively.
Bottom: The blue, green, and orange symbols show the average TKE of the emerging fragments
from TDDFT trajectories starting at the corresponding initial points shown in the top panel.
The red triangles show the average of the particle-number projected TKE distributions obtained from all TDDFT fission trajectories in the top panel. The data are from Ref.~\cite{Hambsch1997NPA}.
}
\label{fig_6}
\end{figure}

As a second example, we therefore consider spontaneous fission of $^{252}$Cf. In spontaneous fission, the system starts in the ground-state potential well, traverses the fission barrier by quantum tunneling, and emerges on the post-barrier side with the energy conserved along the tunneling path. The TDDFT fission trajectories are initialized at exit points beyond the fission barrier, indicated by open circles in the upper panel of Fig.~\ref{fig_6}, where the system emerges from the quantum-tunneling regime. In addition to trajectories initiated from points constrained in $(\beta_{20},\beta_{30})$, and the very asymmetric trajectories starting from points constrained in $(\beta_{20},\beta_{30},\beta_{40})$, we find that closing the gap around $\beta_{30} \approx 1$ requires breaking axial symmetry. The corresponding fission paths are therefore initiated from points constrained in $(\beta_{20},\beta_{30}, \beta_{31})$. Axially symmetric trajectories starting in this region of octupole deformation do not reach scission~\cite{Li2026PRL}.

The resulting TKEs are shown in the lower panel of Fig.~\ref{fig_6}. Open symbols denote the total kinetic energies calculated for the average mass number of the heavy fragment, with the color coding corresponding to the trajectories shown in the upper panel. Although the average TKEs are generally in agreement with the data of Ref.~\cite{Hambsch1997NPA}, and even exceed the experimental values for nearly symmetric fission, their detailed mass dependence is not reproduced. The red triangles again denote the TKEs obtained by averaging over the PNP distributions. In this case, the theoretical results closely follow the empirical trend of the TKE as a function of the heavy-fragment mass number. As in the case of $^{240}$Pu, these results demonstrate that allowing for a broader space of collective degrees of freedom and incorporating PNP for the emerging fission fragments leads to an average over the TKE contributions associated with a given fragment pair that reproduces the mass dependence observed experimentally.

\section{Summary and outlook}\label{sec_sum}

In this work we have examined the total kinetic energy of fission fragments of $^{240}$Pu and $^{252}$Cf within the relativistic TDDFT+BCS framework. As shown in our previous studies \cite{Ren2022PRL,Ren2022a,Li2023PRC_FTTD,Li2025PRC}, and recently confirmed independently with a non-relativistic TDHFB study spanning a much more complete set of fragmentations \cite{Bjelcic2026PRC}, standard microscopic fission dynamics reproduces the measured TKE only near the most probable mass split and underestimates it by 10-20 MeV for both more symmetric and more asymmetric fragmentations. Motivated by the observation in Ref. \cite{Bjelcic2026PRC} that this discrepancy persists essentially unchanged under variation of the Skyrme parametrization and of the pairing strength or its surface-versus-volume character, we have investigated two ingredients that are absent from that standard treatment: restoration of the particle-number symmetry broken by pairing correlations, via particle-number projection (PNP), and an extension of the collective space explored on the way to scission beyond the axially symmetric configurations.

We confirm, using a pairing interaction increased by 10\% relative to the value fixed by the empirical odd-even mass staggering of $^{240}$Pu, that varying the pairing strength alone leaves both the fission paths and the resulting TKE essentially unchanged, in agreement with the conclusion of Ref. \cite{Bjelcic2026PRC} obtained with a different energy density functional and pairing treatment. In contrast, particle-number projection changes the picture qualitatively: whereas an unprojected TDDFT+BCS trajectory yields only a single average TKE value at a single average, non-integer fragment mass, PNP converts each trajectory into a distribution of TKE values over the physical, integer (Z,N) partitions it actually populates, correctly decreasing with the mass number of the heavy fragment as the Coulomb repulsion between the receding fragments decreases. We further find that the initial thickness of the pre-scission neck, controlled by the hexadecapole moment $\beta_{40}$, has a direct and sizable effect on the TKE: constraining a trajectory to start with a thinner neck brings the fragment charge distributions closer together at scission and raises the TKE by nearly 10 MeV, without changing the fragment mass split. Combining both ingredients, an ensemble of trajectories spanning the enlarged
collective space, each projected onto good particle number, and averaged according to which trajectories contribute to a given fragment pair, reproduces the mass dependence of the experimental TKE systematics considerably better than any single, unprojected trajectory, both for induced fission of $^{240}$Pu and for spontaneous fission of $^{252}$Cf.

These results indicate that a part of the long-standing TKE discrepancy in microscopic fission theory is not a deficiency of the energy density functional itself, but reflects the incompleteness of representing the fissioning system by a single, symmetry-violating mean-field trajectory confined to a restricted collective space. Restoring particle-number symmetry and allowing the system to explore a richer set of pre-scission shapes each address a distinct part of the problem. The former corrects the mapping between a mean-field trajectory and the physical, integer Z and N fragments it produces, while the latter corrects the fissioning system's ability to reach the correct scission configuration in the first place. At the same time, our results remain only partially satisfactory. The improvement obtained here is primarily in the mass dependence, or shape, of TKE(A). A systematic offset relative to experiment persists, most visibly for $^{252}$Cf at large heavy-fragment mass. Moreover, for induced fission of $^{240}$Pu the projected and averaged TKE should be understood as a lower bound, since our trajectories start from configurations on the deformation energy surface of the cold nucleus and thus omit the pre-scission excitation energy that, in neutron-induced fission, is only partly retained by the fragments as internal excitation and not entirely as kinetic energy. A consistent comparison with data will require combining the present approach with an explicit treatment of this excitation energy and of its partition between the fissioning system and the emerging fragments.

Several extensions of this work suggest themselves. The choice of collective space considered here, while already broader than in most previous studies, is not exhaustive, and further non-axial and higher-multipole degrees of freedom may be needed to fully capture the relevant scission configurations, particularly in the symmetric and very asymmetric mass regions. The present trajectory-by-trajectory PNP averaging could be replaced by a proper configuration-mixing treatment of the collective coordinates, in the spirit of the time-dependent generator coordinate method. Finally, having shown that these ingredients materially improve the predicted TKE, it will be important to test whether they similarly improve the closely related, and similarly problematic, prediction of how the total excitation energy is shared between the two fission fragments \cite{Bjelcic2026PRC}. Extending the present framework to additional fissioning systems, spanning a wider range of mass and fissility, would help establish how general these conclusions are.

\clearpage
\section{Appendix}\label{sec_app}
In TDDFT, the wave function of a nucleus can be expressed as a direct product of proton and neutron wave functions
\begin{equation}
    |\Phi\rangle = |\Phi_n\rangle \bigotimes  |\Phi_p\rangle = \prod_{i>0}[\mu_{i,n}+\nu_{i,n} c_{i,n}^\dagger c_{\bar{i},n}^\dagger]|-\rangle
    \bigotimes  \prod_{j>0}[\mu_{j,p}+\nu_{j,p}c_{j,p}^\dagger c_{\bar{j},p}^\dagger]|-\rangle,
\end{equation}
where the subscript $n$ ($p$) denotes the parameters related to neutrons (protons).
The particle-number projected state reads
\begin{equation}
 \hat{P}_N\hat{P}_{N_f}^{V_f}\hat{P}_Z\hat{P}_{Z_f}^{V_f}|\Phi\rangle = \hat{P}_N\hat{P}_{N_f}^{V_f} \prod_{i>0}[\mu_{i,n}+\nu_{i,n} c_{i,n}^\dagger c_{\bar{i},n}^\dagger]|-\rangle
    \bigotimes  \hat{P}_Z\hat{P}_{Z_f}^{V_f}\prod_{j>0}[\mu_{j,p}+\nu_{j,p}c_{j,p}^\dagger c_{\bar{j},p}^\dagger]|-\rangle,
\end{equation}
with the explicit expression for $\hat{P}_M\hat{P}_{M_f}^{V_f}\prod_{j>0}[\mu_{j,m}+\nu_{j,m}c_{j,m}^\dagger c_{\bar{j},m}^\dagger]|-\rangle$ ($M = Z,N$ and $m = p,n$):
\begin{equation}
\begin{aligned}
    &\hat{P}_M\hat{P}_{M_f}^{V_f}\prod_{j>0}[\mu_{j,m}+\nu_{j,m}c_{j,m}^\dagger c_{\bar{j},m}^\dagger]|-\rangle\\
    &= \frac{1}{4\pi^2}\int d\theta_f^m \int d\theta^m~e^{-i\theta_f^m M_f-i\theta^m M}~e^{i\theta^m\hat{M}+i\theta_f^m\hat{M}_f}\prod_{j>0}[\mu_{j,m}+\nu_{j,m}c_{j,m}^\dagger c_{\bar{j},m}^\dagger]|-\rangle\\
    &= \frac{1}{4\pi^2}\int d\theta_f^m \int d\theta^m~e^{-i\theta_f^m M_f-i\theta^m M} \prod_{j>0}[\mu_{j,m}+\nu_{j,m} b_{j,m}^\dagger(\theta_f^m,\theta^m) b_{\bar{j},m}^\dagger(\theta_f^m,\theta^m)]|-\rangle .
\end{aligned}
\end{equation}
The operators $b_{j,m}^\dagger(\theta_f^m,\theta^m)$ are defined:
\begin{equation}
b_{j,m}^\dagger(\theta_f^m,\theta^m) = e^{i\theta^m} e^{i\theta_f^m\hat{M}_f} c_{j,m}^{\dagger} e^{-i\theta_f^m\hat{M}_f},
\end{equation}
and satisfy the orthonormality condition,
\begin{equation}
\langle-|b_{i,m}(\theta_f^m,\theta^m) b_{j,m}^\dagger(\theta_f^m,\theta^m)|-\rangle = \langle-|c_{i,m}c_{j,m}^{\dagger}|-\rangle = \delta_{ij}.
\end{equation}
Thus, $|\Phi_m(\theta_f^m,\theta^m)\rangle=\prod_{j>0} [\mu_{j,m}+\nu_{j,m} b_{j,m}^\dagger(\theta_f^m,\theta^m) b_{\bar{j},m}^\dagger(\theta_f^m,\theta^m)]|-\rangle$ is equivalent to a BCS vacuum.
The denominator in Eq.~(\ref{Eq_pj_curr}) factorizes as follows:
\begin{equation}
\langle\Phi|\hat{P}_N\hat{P}_{N_f}^{V_f}\hat{P}_Z\hat{P}_{Z_f}^{V_f}|\Phi\rangle =
\langle\Phi_n|\hat{P}_N\hat{P}_{N_f}^{V_f}|\Phi_n\rangle \langle \Phi_p|\hat{P}_Z\hat{P}_{Z_f}^{V_f}|\Phi_p\rangle,
\end{equation}
where the overlap kernel $\langle\Phi_m|\hat{P}_M\hat{P}_{M_f}^{V_f}|\Phi_m\rangle$ reads ($M = Z,~N$ and $m = p,~n$),
\begin{equation}
\langle\Phi_m|\hat{P}_M\hat{P}_{M_f}^{V_f}|\Phi_m\rangle = \frac{1}{4\pi^2}\int d\theta_f^m \int d\theta^m~e^{-i\theta_f^m M_f-i\theta^m M} \langle \Phi_m|\Phi_m(\theta_f^m,\theta^m)\rangle,
\label{Eq_Proj_M}
\end{equation}
and $\langle \Phi_m|\Phi_m(\theta_f^m,\theta^m)\rangle$ can be computed using the Pfaffian algorithms~\cite{Hu2014PLB,Robledo2009PRC}. Correspondingly, the numerator in Eq.~(\ref{Eq_pj_curr}) can be expressed
\begin{equation}
\begin{aligned}
&\langle\Phi|\bar{\psi} \gamma^\mu \psi \hat{P}_N\hat{P}_{N_f}^{V_f}\hat{P}_Z\hat{P}_{Z_f}^{V_f}|\Phi\rangle\\
&= \langle\Phi_n|\bar{\psi}_n \gamma^\mu \psi_n \hat{P}_N\hat{P}_{N_f}^{V_f}|\Phi_n\rangle \langle \Phi_p|\hat{P}_Z\hat{P}_{Z_f}^{V_f}|\Phi_p\rangle +\langle\Phi_n|\hat{P}_N\hat{P}_{N_f}^{V_f}|\Phi_n\rangle \langle \Phi_p|\bar{\psi}_p \gamma^\mu \psi_p\hat{P}_Z\hat{P}_{Z_f}^{V_f}|\Phi_p\rangle,
\end{aligned}
\end{equation}
where $\psi_n$ and $\psi_p$ are the neutron and proton field operators, respectively.\\
The overlap $\langle\Phi_m|\hat{P}_M\hat{P}_{M_f}^{V_f}|\Phi_m\rangle$ is computed using Eq.~(\ref{Eq_Proj_M}), and
the overlap $\langle\Phi_m|\bar{\psi}_m \gamma^\mu \psi_m \hat{P}_M\hat{P}_{M_f}^{V_f}|\Phi_m\rangle$ ($M = Z,N$ and $m = p,n$) reads
\begin{equation}
\begin{aligned}
& \langle\Phi_m|\bar{\psi}_m \gamma^\mu \psi_m \hat{P}_M\hat{P}_{M_f}^{V_f}|\Phi_m\rangle
= \frac{1}{4\pi^2} \int d\theta_f^m \int d\theta^m~e^{-i\theta_f^m M_f-i\theta^m M} \langle \Phi_m|\bar{\psi}_m \gamma^\mu \psi_m|\Phi_m(\theta_f^m,\theta^m)\rangle\\
& = \frac{1}{4\pi^2} \int d\theta_f^m \int d\theta^m~e^{-i\theta_f^m M_f-i\theta^m M} \sum_{i,j}
\bar{\phi}_{i,m}({\bm r_f}) \gamma^\mu \phi_{j,m}({\bm r_f},\theta_f^m,\theta^m) \langle \Phi_m|c_{i,m}^\dagger b_{j,m}(\theta_f^m,\theta^m)|\Phi_m(\theta_f^m,\theta^m)\rangle,\\
\end{aligned}
\end{equation}
where $\phi_{j,m}({\bm r_f},\theta_f^m,\theta^m)$ is the wave function corresponding to the operator $b_{j,m}^{\dagger}(\theta_f^m,\theta^m)$. The Pfaffian algorithm is used to compute the overlap kernel $\langle \Phi_m|c_{i,m}^\dagger b_{j,m}(\theta_f^m,\theta^m)|\Phi_m(\theta_f^m,\theta^m)\rangle$, and we note that only states in the Fermi sea contribute to the nucleon field operator $\psi_m$ (no-sea approximation),
\begin{equation}
\bar{\psi}_m \approx \sum_i c_{i,m}^\dagger \bar{\phi}_{i,m}({\bm r_f}),~~~\psi_m \approx \sum_j  b_{j,m}(\theta_f^m,\theta^m) \phi_{j,m}({\bm r_f},\theta_f^m,\theta^m).
\end{equation}

\begin{acknowledgments}
This work was supported in part by the High-End Foreign Experts Plan of China,
the National Natural Science Foundation of China (Grants No.12435006 and No.12475117),
the National Key Laboratory of Neutron Science and Technology (Grant No. NST202401016),
the National Key R\&D Program of China 2024YFE0109803,
the National Key Research and Development Program of China 2024YFA1612600,
the National Key Laboratory of Nuclear Data foundation JCKY2026201C153,
the High-Performance Computing Platform of Peking University,
by the project ``Implementation of cutting-edge
research and its application as part of the Scientific Center of Excellence for Quantum
and Complex Systems, and Representations of Lie Algebras", Grant No. PK.1.1.10.0004,
co-financed by the European Union through the European Regional Development Fund -
Competitiveness and Cohesion Programme 2021-2027, and the Croatian Science Foundation project: ``Dynamics of complex femtosystems" (IP-2025-02-4214).
\end{acknowledgments}

\bigskip

%

\end{document}